\documentclass[%
reprint,
 amsmath,amssymb,
 aps,
 prc,
 superscriptaddress,
]{revtex4-2}
\usepackage[utf8]{inputenc}
\usepackage{hyperref} 

\usepackage{dcolumn}
\usepackage{bm}
\usepackage{siunitx}
\usepackage[modulo]{lineno}

\usepackage{pgfplots}
\usepgfplotslibrary{groupplots,dateplot}
\usetikzlibrary{patterns,shapes.arrows}

\pgfplotsset{compat=newest}

\pgfplotsset{error bar legend/.style={%
    /pgfplots/legend image code/.prefix code={%
      \pgfkeysgetvalue{/pgfplots/error bars/error mark}{\pgfplotserrorbarsmark}%
      \draw[%
        /pgfplots/every error bar, 
        mark=\pgfplotserrorbarsmark, 
        /pgfplots/error bars/error mark options, 
        sharp plot,
        ##1
      ] plot coordinates {(0.3cm, -0.15cm) (0.3cm, 0.15cm)};%
    }
  },
  /pgfplots/colormap/inferno high res/.style={%
  /pgfplots/colormap={inferno high res}{
      rgb=(1, 1, 1)
      rgb=(0.001462,  0.000466,  0.013866)
      rgb=(0.002267,  0.00127,  0.01857)
      rgb=(0.003299,  0.002249,  0.024239)
      rgb=(0.004547,  0.003392,  0.030909)
      rgb=(0.006006,  0.004692,  0.038558)
      rgb=(0.007676,  0.006136,  0.046836)
      rgb=(0.009561,  0.007713,  0.055143)
      rgb=(0.011663,  0.009417,  0.06346)
      rgb=(0.013995,  0.011225,  0.071862)
      rgb=(0.016561,  0.013136,  0.080282)
      rgb=(0.019373,  0.015133,  0.088767)
      rgb=(0.022447,  0.017199,  0.097327)
      rgb=(0.025793,  0.019331,  0.10593)
      rgb=(0.029432,  0.021503,  0.114621)
      rgb=(0.033385,  0.023702,  0.123397)
      rgb=(0.037668,  0.025921,  0.132232)
      rgb=(0.042253,  0.028139,  0.141141)
      rgb=(0.046915,  0.030324,  0.150164)
      rgb=(0.051644,  0.032474,  0.159254)
      rgb=(0.056449,  0.034569,  0.168414)
      rgb=(0.06134,  0.03659,  0.177642)
      rgb=(0.066331,  0.038504,  0.186962)
      rgb=(0.071429,  0.040294,  0.196354)
      rgb=(0.076637,  0.041905,  0.205799)
      rgb=(0.081962,  0.043328,  0.215289)
      rgb=(0.087411,  0.044556,  0.224813)
      rgb=(0.09299,  0.045583,  0.234358)
      rgb=(0.098702,  0.046402,  0.243904)
      rgb=(0.104551,  0.047008,  0.25343)
      rgb=(0.110536,  0.047399,  0.262912)
      rgb=(0.116656,  0.047574,  0.272321)
      rgb=(0.122908,  0.047536,  0.281624)
      rgb=(0.129285,  0.047293,  0.290788)
      rgb=(0.135778,  0.046856,  0.299776)
      rgb=(0.142378,  0.046242,  0.308553)
      rgb=(0.149073,  0.045468,  0.317085)
      rgb=(0.15585,  0.044559,  0.325338)
      rgb=(0.162689,  0.043554,  0.333277)
      rgb=(0.169575,  0.042489,  0.340874)
      rgb=(0.176493,  0.041402,  0.348111)
      rgb=(0.183429,  0.040329,  0.354971)
      rgb=(0.190367,  0.039309,  0.361447)
      rgb=(0.197297,  0.0384,  0.367535)
      rgb=(0.204209,  0.037632,  0.373238)
      rgb=(0.211095,  0.03703,  0.378563)
      rgb=(0.217949,  0.036615,  0.383522)
      rgb=(0.224763,  0.036405,  0.388129)
      rgb=(0.231538,  0.036405,  0.3924)
      rgb=(0.238273,  0.036621,  0.396353)
      rgb=(0.244967,  0.037055,  0.400007)
      rgb=(0.25162,  0.037705,  0.403378)
      rgb=(0.258234,  0.038571,  0.406485)
      rgb=(0.26481,  0.039647,  0.409345)
      rgb=(0.271347,  0.040922,  0.411976)
      rgb=(0.27785,  0.042353,  0.414392)
      rgb=(0.284321,  0.043933,  0.416608)
      rgb=(0.290763,  0.045644,  0.418637)
      rgb=(0.297178,  0.04747,  0.420491)
      rgb=(0.303568,  0.049396,  0.422182)
      rgb=(0.309935,  0.051407,  0.423721)
      rgb=(0.316282,  0.05349,  0.425116)
      rgb=(0.32261,  0.055634,  0.426377)
      rgb=(0.328921,  0.057827,  0.427511)
      rgb=(0.335217,  0.06006,  0.428524)
      rgb=(0.3415,  0.062325,  0.429425)
      rgb=(0.347771,  0.064616,  0.430217)
      rgb=(0.354032,  0.066925,  0.430906)
      rgb=(0.360284,  0.069247,  0.431497)
      rgb=(0.366529,  0.071579,  0.431994)
      rgb=(0.372768,  0.073915,  0.4324)
      rgb=(0.379001,  0.076253,  0.432719)
      rgb=(0.385228,  0.078591,  0.432955)
      rgb=(0.391453,  0.080927,  0.433109)
      rgb=(0.397674,  0.083257,  0.433183)
      rgb=(0.403894,  0.08558,  0.433179)
      rgb=(0.410113,  0.087896,  0.433098)
      rgb=(0.416331,  0.090203,  0.432943)
      rgb=(0.422549,  0.092501,  0.432714)
      rgb=(0.428768,  0.09479,  0.432412)
      rgb=(0.434987,  0.097069,  0.432039)
      rgb=(0.441207,  0.099338,  0.431594)
      rgb=(0.447428,  0.101597,  0.43108)
      rgb=(0.453651,  0.103848,  0.430498)
      rgb=(0.459875,  0.106089,  0.429846)
      rgb=(0.4661,  0.108322,  0.429125)
      rgb=(0.472328,  0.110547,  0.428334)
      rgb=(0.478558,  0.112764,  0.427475)
      rgb=(0.484789,  0.114974,  0.426548)
      rgb=(0.491022,  0.117179,  0.425552)
      rgb=(0.497257,  0.119379,  0.424488)
      rgb=(0.503493,  0.121575,  0.423356)
      rgb=(0.50973,  0.123769,  0.422156)
      rgb=(0.515967,  0.12596,  0.420887)
      rgb=(0.522206,  0.12815,  0.419549)
      rgb=(0.528444,  0.130341,  0.418142)
      rgb=(0.534683,  0.132534,  0.416667)
      rgb=(0.54092,  0.134729,  0.415123)
      rgb=(0.547157,  0.136929,  0.413511)
      rgb=(0.553392,  0.139134,  0.411829)
      rgb=(0.559624,  0.141346,  0.410078)
      rgb=(0.565854,  0.143567,  0.408258)
      rgb=(0.572081,  0.145797,  0.406369)
      rgb=(0.578304,  0.148039,  0.404411)
      rgb=(0.584521,  0.150294,  0.402385)
      rgb=(0.590734,  0.152563,  0.40029)
      rgb=(0.59694,  0.154848,  0.398125)
      rgb=(0.603139,  0.157151,  0.395891)
      rgb=(0.60933,  0.159474,  0.393589)
      rgb=(0.615513,  0.161817,  0.391219)
      rgb=(0.621685,  0.164184,  0.388781)
      rgb=(0.627847,  0.166575,  0.386276)
      rgb=(0.633998,  0.168992,  0.383704)
      rgb=(0.640135,  0.171438,  0.381065)
      rgb=(0.64626,  0.173914,  0.378359)
      rgb=(0.652369,  0.176421,  0.375586)
      rgb=(0.658463,  0.178962,  0.372748)
      rgb=(0.66454,  0.181539,  0.369846)
      rgb=(0.670599,  0.184153,  0.366879)
      rgb=(0.676638,  0.186807,  0.363849)
      rgb=(0.682656,  0.189501,  0.360757)
      rgb=(0.688653,  0.192239,  0.357603)
      rgb=(0.694627,  0.195021,  0.354388)
      rgb=(0.700576,  0.197851,  0.351113)
      rgb=(0.7065,  0.200728,  0.347777)
      rgb=(0.712396,  0.203656,  0.344383)
      rgb=(0.718264,  0.206636,  0.340931)
      rgb=(0.724103,  0.20967,  0.337424)
      rgb=(0.729909,  0.212759,  0.333861)
      rgb=(0.735683,  0.215906,  0.330245)
      rgb=(0.741423,  0.219112,  0.326576)
      rgb=(0.747127,  0.222378,  0.322856)
      rgb=(0.752794,  0.225706,  0.319085)
      rgb=(0.758422,  0.229097,  0.315266)
      rgb=(0.76401,  0.232554,  0.311399)
      rgb=(0.769556,  0.236077,  0.307485)
      rgb=(0.775059,  0.239667,  0.303526)
      rgb=(0.780517,  0.243327,  0.299523)
      rgb=(0.785929,  0.247056,  0.295477)
      rgb=(0.791293,  0.250856,  0.29139)
      rgb=(0.796607,  0.254728,  0.287264)
      rgb=(0.801871,  0.258674,  0.283099)
      rgb=(0.807082,  0.262692,  0.278898)
      rgb=(0.812239,  0.266786,  0.274661)
      rgb=(0.817341,  0.270954,  0.27039)
      rgb=(0.822386,  0.275197,  0.266085)
      rgb=(0.827372,  0.279517,  0.26175)
      rgb=(0.832299,  0.283913,  0.257383)
      rgb=(0.837165,  0.288385,  0.252988)
      rgb=(0.841969,  0.292933,  0.248564)
      rgb=(0.846709,  0.297559,  0.244113)
      rgb=(0.851384,  0.30226,  0.239636)
      rgb=(0.855992,  0.307038,  0.235133)
      rgb=(0.860533,  0.311892,  0.230606)
      rgb=(0.865006,  0.316822,  0.226055)
      rgb=(0.869409,  0.321827,  0.221482)
      rgb=(0.873741,  0.326906,  0.216886)
      rgb=(0.878001,  0.33206,  0.212268)
      rgb=(0.882188,  0.337287,  0.207628)
      rgb=(0.886302,  0.342586,  0.202968)
      rgb=(0.890341,  0.347957,  0.198286)
      rgb=(0.894305,  0.353399,  0.193584)
      rgb=(0.898192,  0.358911,  0.18886)
      rgb=(0.902003,  0.364492,  0.184116)
      rgb=(0.905735,  0.37014,  0.17935)
      rgb=(0.90939,  0.375856,  0.174563)
      rgb=(0.912966,  0.381636,  0.169755)
      rgb=(0.916462,  0.387481,  0.164924)
      rgb=(0.919879,  0.393389,  0.16007)
      rgb=(0.923215,  0.399359,  0.155193)
      rgb=(0.92647,  0.405389,  0.150292)
      rgb=(0.929644,  0.411479,  0.145367)
      rgb=(0.932737,  0.417627,  0.140417)
      rgb=(0.935747,  0.423831,  0.13544)
      rgb=(0.938675,  0.430091,  0.130438)
      rgb=(0.941521,  0.436405,  0.125409)
      rgb=(0.944285,  0.442772,  0.120354)
      rgb=(0.946965,  0.449191,  0.115272)
      rgb=(0.949562,  0.45566,  0.110164)
      rgb=(0.952075,  0.462178,  0.105031)
      rgb=(0.954506,  0.468744,  0.099874)
      rgb=(0.956852,  0.475356,  0.094695)
      rgb=(0.959114,  0.482014,  0.089499)
      rgb=(0.961293,  0.488716,  0.084289)
      rgb=(0.963387,  0.495462,  0.079073)
      rgb=(0.965397,  0.502249,  0.073859)
      rgb=(0.967322,  0.509078,  0.068659)
      rgb=(0.969163,  0.515946,  0.063488)
      rgb=(0.970919,  0.522853,  0.058367)
      rgb=(0.97259,  0.529798,  0.053324)
      rgb=(0.974176,  0.53678,  0.048392)
      rgb=(0.975677,  0.543798,  0.043618)
      rgb=(0.977092,  0.55085,  0.03905)
      rgb=(0.978422,  0.557937,  0.034931)
      rgb=(0.979666,  0.565057,  0.031409)
      rgb=(0.980824,  0.572209,  0.028508)
      rgb=(0.981895,  0.579392,  0.02625)
      rgb=(0.982881,  0.586606,  0.024661)
      rgb=(0.983779,  0.593849,  0.02377)
      rgb=(0.984591,  0.601122,  0.023606)
      rgb=(0.985315,  0.608422,  0.024202)
      rgb=(0.985952,  0.61575,  0.025592)
      rgb=(0.986502,  0.623105,  0.027814)
      rgb=(0.986964,  0.630485,  0.030908)
      rgb=(0.987337,  0.63789,  0.034916)
      rgb=(0.987622,  0.64532,  0.039886)
      rgb=(0.987819,  0.652773,  0.045581)
      rgb=(0.987926,  0.66025,  0.05175)
      rgb=(0.987945,  0.667748,  0.058329)
      rgb=(0.987874,  0.675267,  0.065257)
      rgb=(0.987714,  0.682807,  0.072489)
      rgb=(0.987464,  0.690366,  0.07999)
      rgb=(0.987124,  0.697944,  0.087731)
      rgb=(0.986694,  0.70554,  0.095694)
      rgb=(0.986175,  0.713153,  0.103863)
      rgb=(0.985566,  0.720782,  0.112229)
      rgb=(0.984865,  0.728427,  0.120785)
      rgb=(0.984075,  0.736087,  0.129527)
      rgb=(0.983196,  0.743758,  0.138453)
      rgb=(0.982228,  0.751442,  0.147565)
      rgb=(0.981173,  0.759135,  0.156863)
      rgb=(0.980032,  0.766837,  0.166353)
      rgb=(0.978806,  0.774545,  0.176037)
      rgb=(0.977497,  0.782258,  0.185923)
      rgb=(0.976108,  0.789974,  0.196018)
      rgb=(0.974638,  0.797692,  0.206332)
      rgb=(0.973088,  0.805409,  0.216877)
      rgb=(0.971468,  0.813122,  0.227658)
      rgb=(0.969783,  0.820825,  0.238686)
      rgb=(0.968041,  0.828515,  0.249972)
      rgb=(0.966243,  0.836191,  0.261534)
      rgb=(0.964394,  0.843848,  0.273391)
      rgb=(0.962517,  0.851476,  0.285546)
      rgb=(0.960626,  0.859069,  0.29801)
      rgb=(0.95872,  0.866624,  0.31082)
      rgb=(0.956834,  0.874129,  0.323974)
      rgb=(0.954997,  0.881569,  0.337475)
      rgb=(0.953215,  0.888942,  0.351369)
      rgb=(0.951546,  0.896226,  0.365627)
      rgb=(0.950018,  0.903409,  0.380271)
      rgb=(0.948683,  0.910473,  0.395289)
      rgb=(0.947594,  0.917399,  0.410665)
      rgb=(0.946809,  0.924168,  0.426373)
      rgb=(0.946392,  0.930761,  0.442367)
      rgb=(0.946403,  0.937159,  0.458592)
      rgb=(0.946903,  0.943348,  0.47497)
      rgb=(0.947937,  0.949318,  0.491426)
      rgb=(0.949545,  0.955063,  0.50786)
      rgb=(0.95174,  0.960587,  0.524203)
      rgb=(0.954529,  0.965896,  0.540361)
      rgb=(0.957896,  0.971003,  0.556275)
      rgb=(0.961812,  0.975924,  0.571925)
      rgb=(0.966249,  0.980678,  0.587206)
      rgb=(0.971162,  0.985282,  0.602154)
      rgb=(0.976511,  0.989753,  0.61676)
      rgb=(0.982257,  0.994109,  0.631017)
      rgb=(0.988362,  0.998364,  0.644924)
    },
  }
}

\usetikzlibrary{external}
\newcommand\xetre{$^{133}$Xe }
\newcommand\xetres{$^{133}$Xe}
\newcommand\xeto{$^{132}$Xe }
\newcommand\xetos{$^{132}$Xe}

\newcommand{\la}{LaBr$_3$(Ce) }

\newcommand{\gsf}{$\gamma$SF }
\newcommand{\gsfs}{$\gamma$SF}

\newcommand{\degree}{$^\circ$}

\newcommand{\nreaction}{$^{132}$Xe(n,$\gamma$)$^{133}$Xe}

\begin{document}

\title{\texorpdfstring{Statistical properties of $^{133}$Xe and the $^{132}$Xe$(n,
\gamma)$ cross section
}{Statistical properties of 133Xe and the 132Xe(n,
γ) cross section
}}

\author{H.~C.~Berg}\email{bergh@frib.msu.edu}
\affiliation{Physics and astronomy, Michigan State University, East Lansing, 48824, MI USA}
\affiliation{Facility for Rare Isotope Beams, Michigan State University, East Lansing, MI 48824, USA}
\affiliation{Department of Physics, University of Oslo, N-0316 Oslo, Norway}

\author{V.~W.~Ingeberg}\email{vetlewi@fys.uio.no}

\author{S.~Siem}
\affiliation{Department of Physics, University of Oslo, N-0316 Oslo, Norway}
\affiliation{Norwegian Nuclear Research Centre, Norway}

\author{M.~Wiedeking}
\affiliation{Nuclear Science Division, Lawrence Berkeley National Laboratory, Berkeley, CA 94720, USA}
\affiliation{School of Physics, University of the Witwatersrand, 2050 Johannesburg, South Africa}
\affiliation{iThemba LABS, P.O. Box 722, 7129 Somerset West, South Africa}

\author{D.~L.~Bleuel}
\affiliation{Lawrence Livermore National Laboratory, 7000 East Avenue, Livermore, California 94550-9234, USA}
\author{A.~Ratkiewicz} 
\affiliation{Lawrence Livermore National Laboratory, 7000 East Avenue, Livermore, California 94550-9234, USA}

\author{A.~A.~Avaa}\thanks{Current affiliation: TRIUMF, 4004 Wesbrook Mall, Vancouver, V6T 2A3, British Columbia, Canada}
\affiliation{iThemba LABS, P.O. Box 722, 7129 Somerset West, South Africa}

\author{T.~D.~Bucher}
\affiliation{Department of Physics, University  of Cape Town, Rondebosch, 7700, South Africa}
\affiliation{iThemba LABS, P.O. Box 722, 7129 Somerset West, South Africa}

\author{M.~V.~J.~Chisapi}
\affiliation{iThemba LABS, P.O. Box 722, 7129 Somerset West, South Africa}
\affiliation{Department of Pure and Applied Physics, School of Natural and Applied
Sciences, University of Zambia P.O. Box 32379, Lusaka, Zambia}

\author{A.~Görgen}
\affiliation{Department of Physics, University of Oslo, N-0316 Oslo, Norway}
\affiliation{Norwegian Nuclear Research Centre, Norway}
\author{P.~Jones}
\affiliation{iThemba LABS, P.O. Box 722, 7129 Somerset West, South Africa}
\author{B.~V.~Kheswa}
\affiliation{iThemba LABS, P.O. Box 722, 7129 Somerset West, South Africa}
\affiliation{Department of Physics, University of Johannesburg, P.O. Box 524, Auckland Park 2006, South Africa}
\author{K.~L.~Malatji}
\affiliation{iThemba LABS, P.O. Box 722, 7129 Somerset West, South Africa}
\affiliation{Department of Physics, Stellenbosch University, P/B X1, Matieland, 7602, South Africa}
\author{S.~H.~Mthembu}
\affiliation{Department of Physics, University of the Western Cape, P/B X17 Bellville 7535, South Africa}
\affiliation{iThemba LABS, P.O. Box 722, 7129 Somerset West, South Africa}
\author{G.~O'Neill}
\affiliation{Data Management and Software Centre, European Spallation Source ERIC, Copenhagen, Denmark}

\author{P.~Papka}
\affiliation{iThemba LABS, P.O. Box 722, 7129 Somerset West, South Africa}
\affiliation{Department of Physics, Stellenbosch University, P/B X1, Matieland, 7602, South Africa}
\author{L.~Pellegri}
\affiliation{iThemba LABS, P.O. Box 722, 7129 Somerset West, South Africa}
\affiliation{School of Physics, University of the Witwatersrand, Johannesburg 2050, South Africa}
\author{T.~Seakamela}
\affiliation{iThemba LABS, P.O. Box 722, 7129 Somerset West, South Africa}
\author{O.~Shirinda}
\affiliation{Department of Physical and Earth Sciences, Sol Plaatje University, P/B X5008, Kimberley 8301, South Africa}

\author{B.~R.~Zikhali} 
\affiliation{iThemba LABS, P.O. Box 722, 7129 Somerset West, South Africa}
\affiliation{Department of Physics, University of Zululand, P/B X1001, KwaDlangezwa 3886, South Africa}

\date{February 2, 2026}

\begin{abstract}
\begin{description}
\item[Background] 
\xetre is an interesting case for plasma physics to explore nuclear excitation by electron capture, as the process can be studied using statistical properties of \xetres.  
\item[Purpose] In this work we present results on \xetre from the inverse-Oslo method where we extract the nuclear level density and the $\gamma$-strength function, which is used to calculate the (n,$\gamma$) cross section on \xetos. The $\gamma$-strength function of \xetre can constrain the estimated decay rate from nuclear excitation by electron capture.
\item[Method] The $\mathrm{d}(^{132}\mathrm{Xe},\mathrm{p})^{132}\mathrm{Xe}$ reaction was used to create the compound nucleus \xetres, which was recorded with an annular particle telescope and a scintillator array consisting of \la and BGO-shielded HPGe Clover detectors. With the inverse-Oslo method, it is possible to study nuclei that are impossible or unable to manufacture targets from, short lived isotopes, or as in this work, noble gases. 
\item[Results] We present the extracted nuclear level density, and $\gamma$-strength function for \xetres, along with shell-model calculations of the statistical properties of \xetres. These are the first statistical properties extracted below 6 MeV for any xenon isotope.  
\item[Conclusions] We constrain the \xetos(n,$\gamma$)\xetre cross section and reaction rate using the TALYS reaction code.
\end{description}
\end{abstract}

\keywords{inverse kinematics, inverse-Oslo, nuclear level density, Oslo method, gamma strength function}

\maketitle

\section{Introduction} 

One of the most intriguing challenges in modern nuclear physics is the exploration of the origins of the chemical elements in the cosmos. During the burning phases of stars, fusion reactions are mostly responsible to generate nuclei up to iron \cite{Burbidge57}.  The synthesis of heavier species requires processes that occur in more extreme astrophysical environments, where high temperatures and densities initiate and support sequences of neutron capture that lead to the slow  \cite{Kappeler2011}, intermediate \cite{Wiedeking25} and rapid \cite{Cow21} neutron capture processes, and the photodisintegration \cite{Rau13} process. 

For the case of $^{133}$Xe isotope, establishing high‐-precision nuclear data is desirable for its role as a ubiquitous background isotope in nuclear-‐test monitoring. As a decay product of fission in underground or atmospheric detonations, $^{133}$Xe constitutes a persistent signature that needs to be accurately quantified to distinguish treaty-‐relevant releases from routine radiogenic emissions \cite{Achim2016,Goodwin}. The nuclear data of $^{133}$Xe, particularly also the quasi--continuum region must be understood. In this excitation regime, a large number of unresolved levels contribute to statistical $\gamma$‐-ray emission, and hence to the $\gamma$–-ray strength function ($\gamma$SF) and nuclear level density (NLD) which are used in reaction modeling. Recent experiments at the National Ignition Facility have been conducted to measure the $(n,\gamma)$ cross section on $^{132}$Xe in high-‐temperature plasma environments similar in concept to that outlined by Ref.~\cite{Bleuel2016}. The interpretation of those results will benefit from reliable quasi--continuum data on $^{133}$Xe. 
Limited information on experimental neutron capture cross sections on $^{132}$Xe have primarily been measured at thermal neutron energies \cite{exfor}. Data for higher neutron energies are  highly desirable as these will better constrain nuclear reaction models and reduce extrapolation uncertainties. 

When direct measurements of the neutron--capture cross section on nuclei such as 
$^{132}$Xe are challenging or impractical due to the difficulty of fabricating a contaminant--free gas target, the Hauser–-Feshbach \cite{HF} statistical reaction model is routinely used to predict capture rates across a wide energy range. In this framework, the nucleus is assumed to form a compound system whose formation and decay are governed by its statistical properties. Key ingredients to the model are the NLD and $\gamma$SF \cite{Wiedeking2024} and with those the Hauser--Feshbach approach has been shown to yield reliable capture cross‐-section predictions \cite{Laplace2016, Larsen2016, Kheswa2017, Malatji2019, Ingeberg2020}. 

The $\gamma$SF is dominated by the giant electric dipole resonance (GDR) but can also  exhibits additional structures on its low–-energy tail, including scissors \cite{Heyde2010} and pygmy \cite{Lanza2019} resonances as well as a low--energy enhancement (LEE) \cite{Voinov2004}, which has been suggested to be connected to the scissors resonance \cite{Schwengner2017, Chen2025} and associated with a new type of collective motion, the scissors rotation \cite{Chen2025}. This mode has been first observed in $^{56}$Fe \cite{Voinov2004} and subsequently confirmed by an independent method \cite{Wiedeking2012}, and repeatedly measured in various nuclei, see for example Refs.~\cite{Larsen2017, Renstrom2018, Jones2018, Ingeberg2025}. This enhancement can leads to increased $(n,\gamma)$ cross sections for neutron--rich isotopes \cite{Larsen2010, Midtbo2018}. 

Nuclei with long‐lived isomeric states, as found in some xenon isotopes, play an important role in advancing our understanding of the excitation mechanisms of NEEC. These metastable states can provide an experimentally accessible window into the coupling between electronic and nuclear degrees of freedom, enabling investigations of energy transfer pathways. 

Stellar plasmas exhibit complex properties with ultrahigh temperatures, intense radiation fields, and dynamic density profiles. In these hot plasma environments, nuclei not only collide with free nucleons and light particles but also interact with electrons, leading to nuclear plasma interactions (NPI) which have the potential to modify reaction pathways \cite{Aliotta22}. Nuclear excitation by electron capture (NEEC) may occur where a free electron is captured into an atomic orbital while simultaneously depositing its binding energy into the nucleus, promoting it to an excited state. Understanding and quantifying such plasma--induced effects is therefore essential to fully unravel the cosmic origin of the elements. Theoretical calculations have shown that NEEC may significantly enhance the decay of isomeric states, see e.g. Refs.~\cite{Morel2010, Gunst2014, Rzad2023, Zhao2024}. The potential impacts of NEEC in astrophysical environments has also been investigated for fusion reactions \cite{Lee2023} and neutron--capture reactions \cite{Helmrich2014}.

In this work, the NLD and $\gamma$SF of $^{133}\mathrm{Xe}$ have been measured for the first time. These measurements aim to provide optimally constrained statistical properties for analyzing the aforementioned NIF experiment. Additionally, they will help in obtaining an experimentally constrained neutron-–capture cross sections on $^{132}\mathrm{Xe}$ at energies beyond the thermal regime.

In Section~\ref{sec:exp}, we present the experimental setup, and in Section \ref{sec:analysis} the analysis methods. Section~\ref{sec:bayes} reports on the normalization and the experimentally determined NLD and $\gamma$SF of $^{133}\mathrm{Xe}$. Supporting large‑scale shell‑-model calculations and discussion is provided in Section~\ref{sec:shellmodel}, then we give the $^{132}\mathrm{Xe}(n,\gamma)$ cross sections and reaction rates in Section \ref{sec:talys}, and in Section~\ref{sec:conc} we provide concluding remarks.

\section{Experiment}\label{sec:exp} 

The experiment, with the inverse-kinematic $d(^{132}\mathrm{Xe},p)$ reaction, was carried out at iThemba LABS \cite{Azaiez01102020} using the Separated Sector Cyclotron (SSC). In inverse kinematics \cite{Ingeberg2020}, a beam of $^{132}\mathrm{Xe}$ was accelerated to $530$ MeV and directed onto deuterated polyethylene (CD$_2$) targets ($\approx 99\%$ $\mathrm{D}$-enriched) with thicknesses ranging between $0.8$ and $1$ mg/cm$^2$ \cite{Kheswa_2020}). Beam intensity was maintained at $\approx0.5$ pnA over a total running time of $\approx 90$ hours.

Charged particles were detected by an annular \(\Delta E\)-\(E\) telescope positioned 30 mm downstream of the target, covering laboratory angles \(19^\circ\) to \(47^\circ\).  Both the \(\Delta E\) and \(E\)  silicon detectors were 1000 $\mu$m thick, with a 10 $\mu$m thick Al foil upstream to suppress \(\delta\)-electrons and scattered beam particles. Each double--sided S2--type silicon detector \cite{s2} of the telescope was segmented into 48 rings and 16 sectors; however, only the ring signals were recorded for the $\Delta E$ detector. For $\gamma$--ray spectroscopy, six large--volume LaBr$_3$:Ce detectors (3.5 in x 8.0 in) and eight Compton--suppressed high--purity germanium (HPGe) Clover detectors were deployed.  Due to the lower HPGe efficiency at high--$\gamma$ energies and the resulting sparse population of the particle–-$\gamma$ coincidence matrix, the focus of this work is on $\gamma$-rays detected in the LaBr$_3$:(Ce) detectors only.

Energy calibrations were performed before and after the run. The $\gamma$-ray calibration used $^{60}\mathrm{Co}$, $^{152}\mathrm{Eu}$, and $^{56}\mathrm{Co}$ sources and an $\mathrm{AmBe+^\text{nat}Fe}$ source. The $\Delta E$--$E$ telescope was calibrated with a $^{226}\mathrm{Ra}$ $\alpha$--emitting source. 

Pulses from the detectors were acquired with the Pixie-16 system from XIA LLC and events were built and sorted offline. Prompt $\gamma$--rays from LaBr$_3$:Ce in coincidence with particles were selected by placing a $\sim 70$~ns wide gate on the particle--$\gamma$ time spectrum. Background events are found by placing similar time gates around the peaks in the time spectrum related to beam pulses before and after the prompt time peak. The background spectra were appropriately scaled and subtracted from the prompt coincidence spectra.

The \(\gamma\)--ray energies have been corrected for Doppler effects due to the large Lorentz boost of the outgoing $^{133}$Xe. The measured protons exhibit a significant kinematic angular dependence across individual rings which limits the proton--energy resolution to $\approx$ 1 MeV in addition to experimental binning of 300 keV, discussed below, rendering any small nonlinearity in the LaBr\(_3\):(Ce) detector response at high energies negligible. Punch--through events in the \(\Delta E\)–-\(E\) telescope were identified by their characteristic energy losses and corrected via kinematic energy conservation, yielding a $\approx$ 15 \% increase in statistics up to an excitation energy \(E_x=4\) MeV.  Ambiguous events overlapping stopped protons and punch--through protons were excluded from the analysis.  

\begin{figure*}
    \centering
    \includegraphics[width=\linewidth]{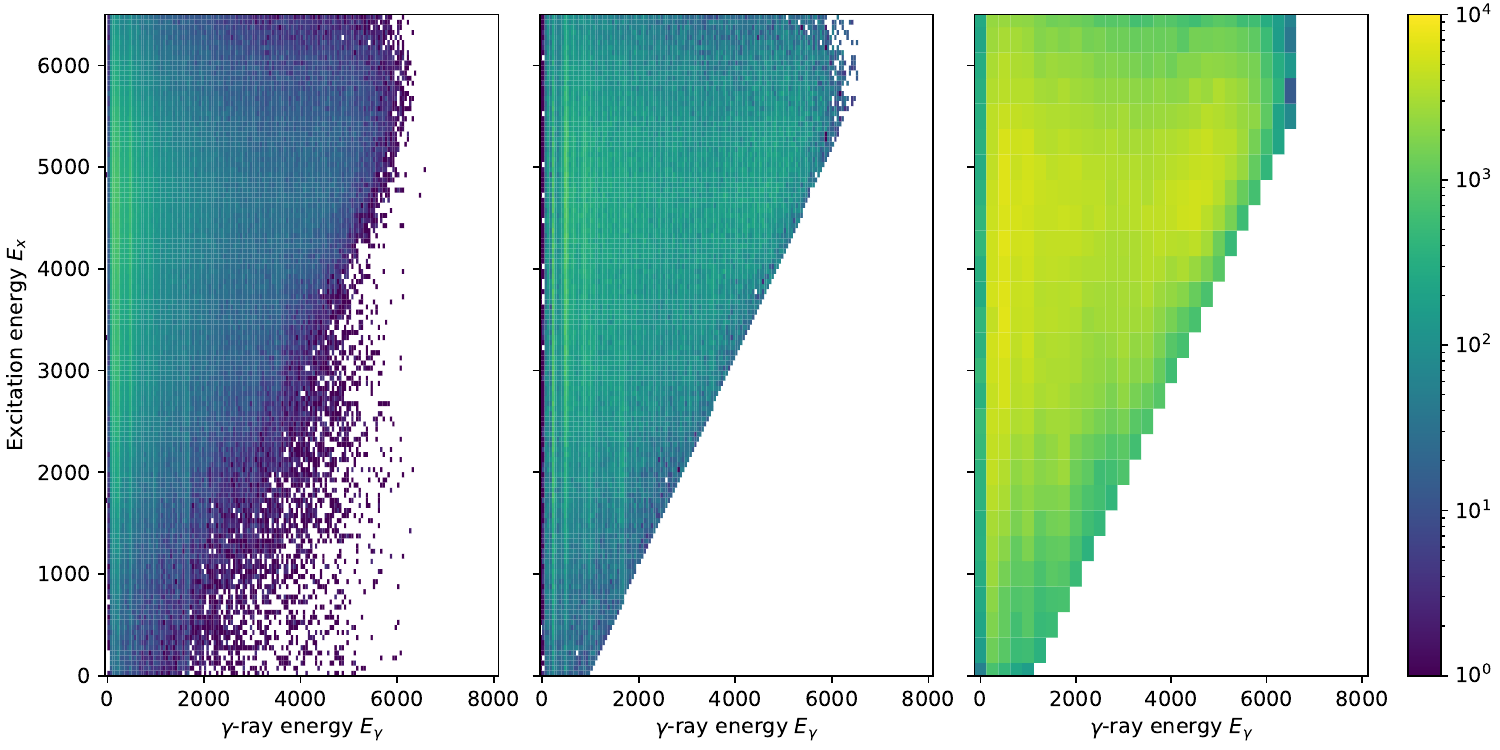}
    \caption{(a) Raw, unprocessed excitation--$\gamma$ coincidence matrix and (b) unfolded matrix with $100$ keV/bins (both axes). The first--generation matrix (c) was re-binned from $100$ to $300$ keV to reduce possible fluctuations. 
    }
    \label{fig:matrix_all}
\end{figure*}

The populated excitation energy for each proton has been reconstructed  by measuring the energy of the outgoing proton in the \(\Delta E\)–-\(E\) telescope. The background subtracted excitation energy versus $\gamma$--ray energy matrix is shown in Fig.~\ref{fig:matrix_all}(a).

\section{Analysis}\label{sec:analysis}
The first step of the Oslo method is the unfolding \cite{GUTTORMSEN1996371} of the particle–-\(\gamma\) coincidence matrices using the detector response function, which was simulated \cite{vetle_w_ingeberg_vetlewiafrodite_2024} using GEANT4 \cite{AGOSTINELLI2003,Allison2006} as this was a new configuration. The raw coincidence matrix (Figs.~\ref{fig:matrix_all}(a)) is first unfolded separately for every \(\gamma\)--detection angle (45\degree, 90\degree, and 135\degree) and the Doppler corrections are included in the unfolding procedure. The same angle--by--angle unfolding procedure is carried out on all background matrices before they are subtracted from the unfolded matrix in Fig.~\ref{fig:matrix_all}(b). 

In the second step the first‐-generation (primary) $\gamma$ spectra are isolated via an iterative subtraction technique of all lower \(E_x\) contributions in the unfolded coincidence matrix \cite{GUTTORMSEN1987518}. The \(E_x\)--\(E_\gamma\) matrix, which is binned in $300$-keV intervals is shown in Fig.~\ref{fig:matrix_all}(c).  
To ensure selection of true primary transitions from the quasi-‐continuum, only \(\gamma\) rays with E$_\gamma \ge 1.8\ \mathrm{MeV}$ and excitation energies  in the range $3.9 \le E_x \le 5.7\ \mathrm{MeV}$ were considered.

The detected \(\gamma\) rays originate in prompt cascades and the probability to emit a first-‐generation \(\gamma\)--ray of energy \(E_\gamma\) from an initial excitation energy \(E_x\) is proportional to the product of the level density and the transmission coefficient \cite{Schiller00},
\begin{equation}
P(E_x,E_\gamma) \propto \rho(E_x-E_\gamma)\mathcal{T}(E_\gamma),
\label{eq:PEgamma}
\end{equation}
\noindent where $\rho(E_x-E_\gamma)$ is the level density fed by the primary transitions and $\mathcal{T}(E_\gamma)$ is the $\gamma$-ray transmission coefficient (assuming the validity of the Brink--Axel Hypothesis \cite{Brink1955, Axel1962}).

From the primary matrix \(P(E_x,E_\gamma)\), we simultaneously extract the nuclear level density (NLD), and the transmission coefficient \(\mathcal{T}(E_\gamma)\) using the $\chi^2$ minimization procedure described in \cite{Schiller00}.

From the first‐-generation matrix \(P(E_x,E_\gamma)\), only the relative functional forms of the NLD and the transmission coefficients are obtained, and they are invariant under transformation \cite{Schiller00}:
\begin{equation}
\begin{split}
    \widetilde{\rho} 
    \left(E_f\right) &= A \rho\left(E_f\right) e^{\alpha E_f} \\
    \widetilde{\mathcal{T}}(E_\gamma) &= B \mathcal{T}\left(E_\gamma\right) e^{\alpha E_\gamma},
\end{split} \label{eq:TransfNLD_GSF}
\end{equation}
\noindent where $A$, $B$, and $\alpha$ are scaling factors, which needs to be determined to get the physical solution. 
The scaling factors are constrained by normalization to auxiliary nuclear data, as described in the next section. 

Assuming that dipole transitions dominate, the transmission coefficient can be related to the $\gamma$SF as
\begin{equation}
\mathcal{T}(E_\gamma) = 2 \pi f(E_{\gamma}) E_{\gamma}^3.
\end{equation}

\noindent The total experimental \(\gamma\)-strength function is expressed as
\begin{equation}
  f(E_{\gamma})
  \approx f_{E1}(E_{\gamma}) + f_{M1}(E_{\gamma}).
\end{equation}
With the current experimental setup we cannot distinguish between the $E1$ and $M1$ strength. A detailed analysis of the systematic uncertainties related to the extraction of NLD and $\mathcal{T}$ is discussed in Ref.~\cite{Larsen2011}.

\section{Bayesian inference of normalization parameters}\label{sec:bayes}
To determine the normalization parameters for the NLD and $\gamma$SF we employ \textit{Bayesian inference}, which is a statistical technique based on Bayes' theorem. Applied to Oslo method data, the theorem can be formulated as \cite{Ingeberg2022, MIDTBO2021}
\begin{equation}
    P(\bm{\theta}|\{ \rho_i \}, \{ f_i \}) = \frac{\mathcal{L}(\{ \rho_i \}, \{ f_i \} | \bm{\theta}) P(\bm{\theta}) }{P(\{ \rho_i \}, \{ f_i \})}, \label{eq:BayesTheorem}
\end{equation}
where $P(\bm{\theta}|\{ \rho_i \}, \{ f_i \})$ is the posterior probability distribution for the set of normalization parameters $\bm{\theta}$,
given un-normalized level density ($\rho$) and $\gamma$-ray strength function ($f$). The likelihood $\mathcal{L}(\{ \rho_i \}, \{ f_i \} | \bm{\theta})$ and prior probability $P(\bm{\theta})$ are discussed in Sec. \ref{sec:Likelihood} and Sec. \ref{sec:Prior}, respectively. The evidence $P(\{ \rho_i \}, \{ f_i \})$ represents the measurement probability of the un-normalized NLD and $\gamma$SF, and can be treated as a normalization factor. Simultaneously with the inference of scaling factors $A$, $B$ and $\alpha$, a parameterization of the $\gamma$SF are also found.
The inference of the normalization parameters ($\bf{\theta}$) is performed by the \verb+ULTRANEST+ package \cite{2021JOSS6.3001B} using the multimodal nested sampling technique \cite{Feroz2008,Feroz2009,Feroz2019}.

\subsection{Likelihood\label{sec:Likelihood}}
The likelihood function $\mathcal{L}(\{ \rho_i \}, \{ f_i \} | \bm{\theta})$ is the conditional probability of measuring the set of un-normalized NLD $\{ \rho_i \}$ and $\gamma$SF $\{ f_i \}$ given the set of normalization and model parameters $\bm{\theta}$. The un-normalized NLD and $\gamma$SF values are assumed to be Gaussian. The total likelihood function can be expressed as the product
\begin{equation}
    \begin{split}
    &\mathcal{L}(\{ \rho_i \}, \{ f_i \} | \bm{\theta}) = \\
     &\mathcal{L}_\text{discrete}(\bm{\theta})\mathcal{L}_\text{CT}(\bm{\theta})\mathcal{L}_{\rho(S_n)}(\bm{\theta})\mathcal{L}_{\langle\Gamma_{\gamma0}\rangle}(\bm{\theta}).
    \end{split}
    \label{eq:LikelihoodTot}
\end{equation}
The likelihood
\begin{equation}
\begin{split}
    \ln\mathcal{L}_\text{discrete}(\bm{\theta}) = \sum_j \frac{1}{\sqrt{2\pi A\sigma_{\rho_j}e^{\alpha E_j}}} \\
    - \frac{1}{2}\sum_j\left( \frac{\rho_{\text{discrete},j} - A\rho_je^{\alpha E_j}}{A\sigma_{\rho_j}e^{\alpha E_j}} \right)^2
\end{split} \label{eq:LikeDisc}
\end{equation}
where the sum runs over excitation energies between $0 \leq E_x \leq 0.8$ MeV and $\rho_\text{discrete}$ are calculated from the known tabulated levels \cite{Feroz2008}.  
A constant temperature model is used to interpolate between the energy region where the experimental NLD are extracted and the level density at the neutron separation energy $\rho(S_n)$, giving rise to the likelihood term
\begin{equation}
    \begin{split}
        \ln\mathcal{L}_\text{CT}(\bm{\theta}) = \sum_j \frac{1}{\sqrt{2\pi A\sigma_{\rho_j}e^{\alpha E_j}}} \\
        - \frac{1}{2}\sum_j\left( \frac{\rho_\text{CT}(E_j) - A\rho_je^{\alpha E_j}}{A\sigma_{\rho_j}e^{\alpha E_j}} \right)^2
    \end{split} \label{eq:LikeCT}
\end{equation}
where $\rho$ and $\sigma_\rho$ are the experimental un-normalized NLD and the standard deviation, respectively.
The constant temperature model is \cite{ERICSON1959481}
\begin{equation}
    \rho_\text{CT}(E) = \frac{1}{T}\exp\left(\frac{E - \delta_0}{T}\right),
\end{equation}
where $\delta_0$ is the shift parameter and $T$ is the temperature. The fitting region for the CT model to the level density is $2.5 \leq E_x \leq 3.5$ MeV.

Assuming equiparity, the NLD at $S_n$ is found from the average neutron resonance spacing of s-wave resonances
\begin{equation}
    \rho(S_n) = \frac{2}{D_0} \frac{1}{g(S_n,J_t + 1/2) + g(S_n, J_t - 1/2)}. \label{eq:RhoSn}
\end{equation}
The $D_0$ is the average s-wave resonance spacing, $J_t$ is the ground state spin of the $A-1$ nucleus, and
\begin{equation}
    g(E_x,J) = \exp\left(-\frac{J^2}{2\sigma^2(E_x)}\right) - \exp\left(-\frac{(J+1)^2}{2\sigma^2(E_x)}\right) \label{eq:SpinDist}
\end{equation}
is the spin distribution \cite{PhysRev.50.332,doi:10.1080/00018736000101239}.
The spin-cutoff parameter is parameterized as \cite{PhysRevC.96.024313}
\begin{equation}
\sigma^2(E_x) = \begin{cases}
    \sigma_d^2 & \quad E < E_d \\
    \sigma_d^2 + \frac{E - E_d}{S_n - E_d}\left(\sigma^2_{S_n} - \sigma^2_d\right) & \quad E \geq E_d.
    \end{cases}
    \label{eq:SpinCutParam}
\end{equation}
where $\sigma_d$ is the spin cutoff at the discrete level ($E_d=2.0$ MeV), estimated from known levels \cite{NDS_A133,Vogt2017,Kaya2018}, and $\sigma_{S_n}$ is at the neutron separation energy. 
This leads to the likelihood term
\begin{equation}
    \begin{split}
        \ln\mathcal{L}_{\rho(S_n)}(\bm{\theta}) = \frac{1}{\sqrt{2\pi \sigma_{\rho(S_n)}}} \\
        - \frac{1}{2}\left( \frac{\exp\left(\frac{\left(S_n-\delta_0\right)}{T}\right)/T - \rho(S_n)}{\sigma_{\rho(S_n)}} \right)^2
    \end{split} \label{eq:LikeRhoSn}
\end{equation}
for NLD at the neutron separation energy. Lastly, the magnitude of the \gsf is determined through the average radiative width, $\langle \Gamma_{\gamma 0} \rangle$ of s-wave resonances, which can be calculated by \cite{PhysRevC.41.1941}:
\begin{equation}\label{eq:Bnormalization}
\begin{split}
\langle \Gamma_{\gamma}(S_n,\bm{\theta})\rangle =
\frac{D_0\,B}{2\pi}
  \int_{0}^{S_n} dE_\gamma
    \mathcal{T}(E_{\gamma})\,\rho(S_n - E_{\gamma}) \\
    \times (g(S_n - E_{\gamma},J_t+1/2) + g(S_n - E_{\gamma},J_t-1/2)).
\end{split}
\end{equation}
This gives rise to the likelihood term
\begin{equation}
    \begin{split}
        \ln\mathcal{L}_{\langle \Gamma_{\gamma0}\rangle}(\bm{\theta}) = \frac{1}{\sqrt{2\pi\sigma_{\left\langle\Gamma_{\gamma0}\right\rangle}}} \\
        - \frac{1}{2}\left( \frac{\langle \Gamma_{\gamma}(S_n,\bm{\theta})\rangle - \langle\Gamma_{\gamma0}\rangle}{\sigma_{\langle\Gamma_{\gamma0}\rangle}} \right).
    \end{split}
\end{equation}

\begin{table}[htbp]
    \centering
    \caption{Values for the spin-cutoff parameter, $\sigma(S_n)$ with the corresponding level density, $\rho(S_n)$ at neutron separation energy for different models.}
    \begin{tabular}{c|c|c}\hline\hline
         Model & $\sigma(S_n)$ & $\rho(S_n)$ $\left(\mathrm{MeV}^{-1}\right)$ \\ \hline
         RMI        & 5.610 & 7.8015$\times10^4$ \\ 
         G\&C FG    & 4.674 & 9.3635$\times10^4$ \\
         E\&B CT    & 4.047 & 3.1764$\times10^4$ \\
         E\&B FG    & 4.348 & 2.9243$\times10^4$ \\ \hline\hline
    \end{tabular}
    \label{tab:sig_Sn}
\end{table}

\begin{figure}[htbp]
    \centering
    \includegraphics[width=0.5\textwidth]{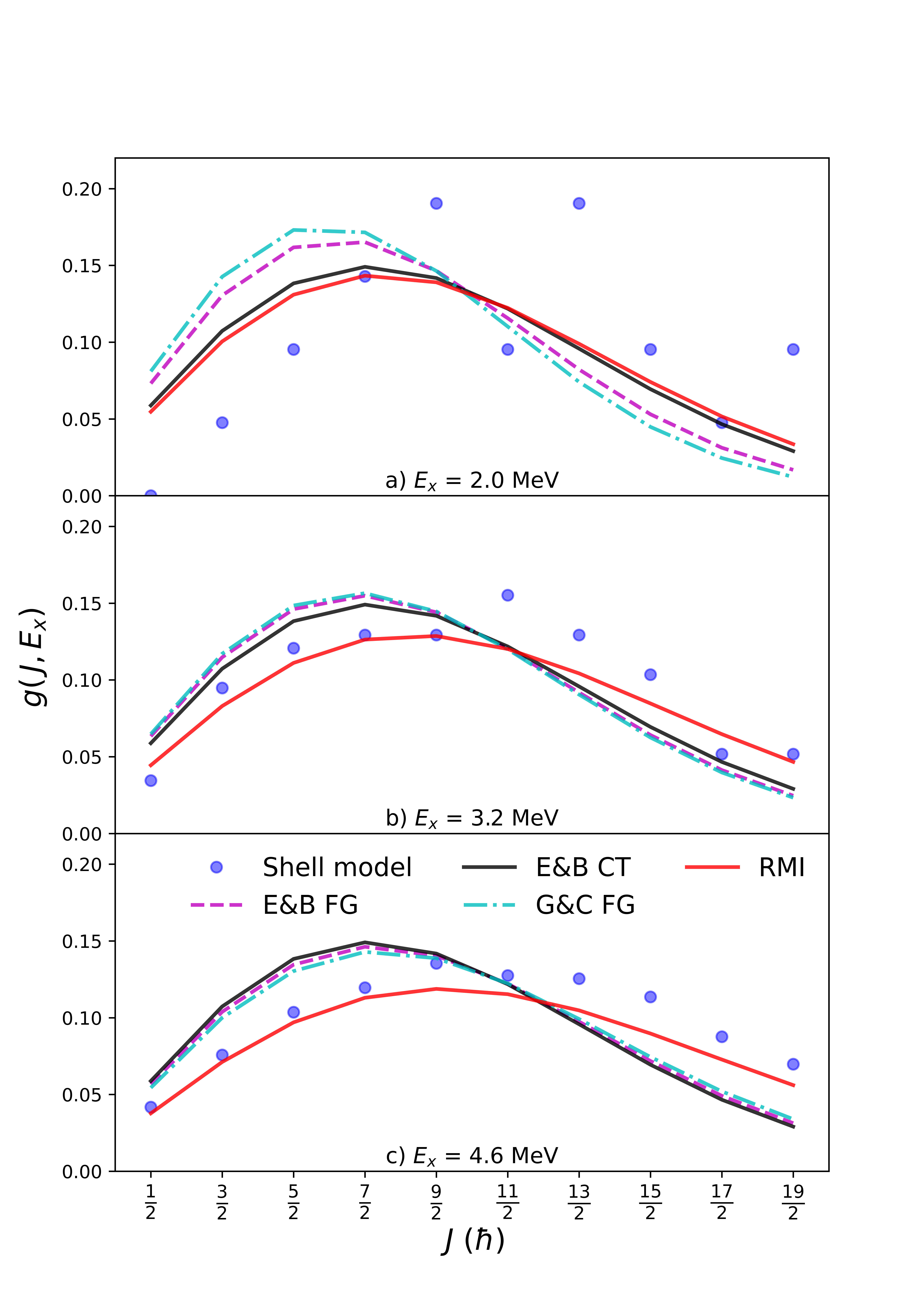}
    \caption{Model dependent spin distribution, $g(J,E_x)$ plotted together with the shell model spin distribution (blue dots). The solid red line is the rigid moment of inertia (RMI) with a reduction factor of 0.8, the solid black line is the constant temperature model with parameters from Egidy and Bucurescu (E\&B CT), the dashed magenta line is the Fermi gas model with parameters from Egidy and Bucurescu (E\&B FG) \cite{eb2006,ebe2006}, and the dash-dotted cyan line is the Fermi gas model with parameters from Gilbert and Cameron (G\&C FG) \cite{Gilbert1965}. (a) $g(J, E_x = 2\,\mathrm{ MeV})$, (b) $g (J, E_x = 3.2\,\mathrm{ MeV})$, (c) $g (J, E_x = 4.6\,\mathrm{ MeV})$.}
    \label{fig:SM_spin_dist}
\end{figure}

Model dependent spin-distributions are plotted together with the shell model spin-distribution in Fig.~\ref{fig:SM_spin_dist}. As a general comparison of the model dependent spin--cutoff parameter at the neutron separation energy, $\sigma_{\rho(S_n)}$, these values are given in Table~\ref{tab:sig_Sn}. Based on this comparison, the RMI spin-distribution with a reduction factor of 0.8 reproduces the shell model calculations the best, as discussed in Section \ref{sec:shellmodel}. With the similarity in our extracted nuclear level density and the shell model level density, we choose the spin-cut off as $\sigma^2_{\mathrm{RMI}} = 5.610$.

The auxiliary nuclear data parameters ($D_0$, $\langle \Gamma_{\gamma0}\rangle$) used in the normalization of NLD and $\gamma$SF are summarized in Table~\ref{tab:OM_prop}.

\begin{table}[htbp]
	\centering
\caption{The parameters used in the normalization procedure \(\langle\Gamma_{\gamma}(S_n)\rangle\), \(D_0\), and \(a\). }
	\label{tab:OM_prop}
\begin{tabular}{lll}\hline \hline
		Parameter & Value & Ref.\\ \hline
		$S_n$ & $6.435$ MeV & \cite{KHAZOV2011855} \\
		$D_0$ &  $750 \pm 230$ eV & \cite{CAPOTE20093107}\\
		$\langle \Gamma_{\gamma 0} \rangle$ & $70 \pm 10$ meV & \cite{saidmughabghab2018}\\  \hline
	\end{tabular}
\end{table}

\subsection{Prior\label{sec:Prior}}
The prior probability distribution is
\begin{equation}
    P(\bm{\theta}) = \prod_{i\in\bm{\theta}} P(\theta_i) \label{eq:Prior}
\end{equation}
where $i$ runs over all parameters in $\bm{\theta} = (A, B, \alpha, T, \delta_0, \sigma_D, \sigma_{S_n})$.
For the normalization parameters $A$, $B$ and $\alpha$ there are no prior information available as there are an infinite number of possible solutions for the extracted NLD and $\gamma$SF points. Ideally the prior probability distribution for these parameters should be an infinite uniform distribution, however with such a prior it is computationally difficult. To address this issue their priors are modeled as normal distributions, with $A$ and $B$ truncated at $0$ since negative value are not physical. The centroid of the prior distributions for the normalization parameters $A$, $B$ and $\alpha$ are taken as the maximum likelihood estimator (MLE), found by maximizing Eq. \eqref{eq:LikelihoodTot}. The variance/width of the prior distributions are set to $5$, $10$ times the MLE for $A$ and $B$, respectively, and $1$ MeV$^{-1}$ for $\alpha$.

The prior for the model parameters of the NLD are assumed to follow a normal distribution, with the temperature parameter $T$ truncated at $0$ MeV to avoid un-physical values. For both the temperature and the shift parameter the centroid of the prior probability distribution, as in the case of the scaling parameters, was taken as the MLE and a width of $2$ MeV and $10$ MeV for temperature and the shift parameter, respectively. The spin-cutoff for the discrete values is estimated from all tabulated states up to $2$ MeV, while the spin-cutoff parameter was taken from the Rigid Moment of Inertia (RMI) model \cite{VonEgidy2005,PhysRevC.73.049901} with a 0.8 reduction factor. This choice was justified from shell-model calculations, see Sect. \ref{sec:shellmodel}. For both $\sigma_D$ and $\sigma_{\rho_{S_n}}$ an uncertainty of $10\%$ was assumed. Numerical values for the probability distributions of the NLD model parameters are given in Table~\ref{tab:Prior_params}.

\begin{table}
\centering
\caption{\label{tab:Prior_params}Parameters for the prior probability distribution for NLD model parameters.}
\begin{tabular}{c|cc} \hline \hline
  Parameter   & $\mu$ & $\sigma$ \\ \hline
  Temperature ($T$) & $0.63$ MeV & $2$ MeV  \\
  Shift ($\delta_0$) & $+0.22$ MeV & $10$ MeV \\
  Discrete spin-cutoff ($\sigma_D$) & $2.30$ & $0.23$ \\
  Spin-cutoff at $S_n$ ($\sigma_{S_n}$) & $5.95$ & $0.60$ \\ \hline
\end{tabular}
\end{table}

\subsection{Results}
\begin{figure}[htbp]
    \centering
    \includegraphics[width=\linewidth]{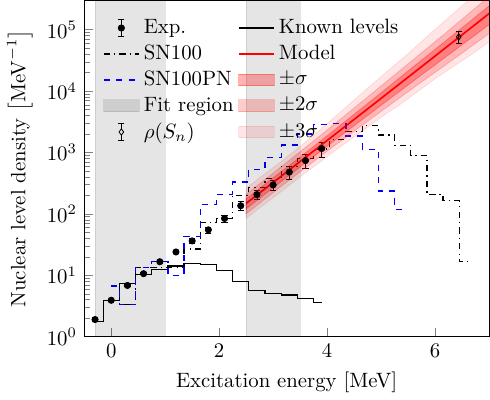}
    \caption{Experimental NLD for \xetres{} (filled circles), the NLD at $S_n$ (solid diamond), known low-lying levels \cite{CAPOTE20093107} (solid black line), the NLD from large-scale shell model calculations with the SN100 and SN100PN (dashed-dotted and dashed lines, respectively), and the extrapolated NLD from the Constant Temperature model (solid blue line). The red shaded bands indicate the credibility intervals for $\sigma$, $2\sigma$ and $3\sigma$ of the model interpolation. The shaded gray areas indicate the energy regions to which the experimental NLD was normalized to discrete level density and the CT model.}
    \label{fig:nld}
\end{figure}

\begin{figure}[htbp]
    \centering
    \includegraphics[width=\linewidth]{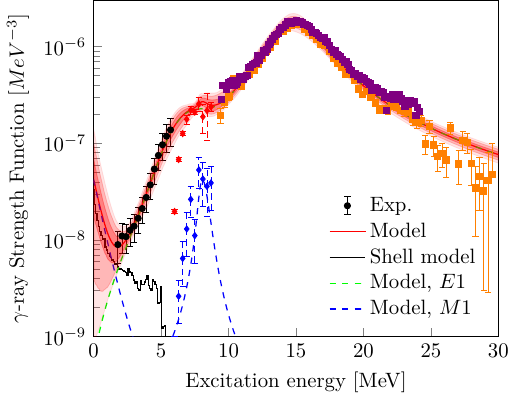}
    \caption{The experimental \gsf of \xetres{} shown with the filled black circles while the red and blue diamonds shows the nuclear resonance fluorescence results in $^{132}\mathrm{Xe}$ and $^{134}\mathrm{Xe}$, respectively \cite{Massarczyk2014}. The $M1$ strength function from large--scale shell model calculations are shown by the black line, see Sect. \ref{sec:shellmodel} for more details.}
    
    \label{fig:gsf_zoomed}
\end{figure}

For each sampled set of parameters $\bm{\theta}_i$ the normalized NLD and $\gamma$SF is calculated, and for each NLD and $\gamma$SF point the marginal posterior distribution is found. The resulting NLD and $\gamma$SF are shown in Figs. \ref{fig:nld} and \ref{fig:gsf_zoomed}, respectively. The resulting model parameters from the marginal posterior distribution is presented in Table~\ref{tab:OM_param_post_NLD}. 

\begin{table}[htbp]
    \centering
    \caption{Posterior mean of normalization and NLD model parameters.\label{tab:OM_param_post_NLD}}
    \begin{tabular}{c|c}\hline \hline
        Parameter & Value \\ \hline
        $T$ &  $0.634(30)$ MeV \\ 
        $\delta_0$ & $-0.36(22)$ MeV \\ \hline
        $\sigma_D$ & $2.34(23)$  \\ 
		$\sigma_{\rho(S_n)}$ & $5.45(60)$ \\ \hline \hline
    \end{tabular}
\end{table}

\subsection{\texorpdfstring{Parameterization of the $\gamma$SF}{Parameterization of the gSF}}
To parameterize the $\gamma$SF, the $E1$-strength component is assumed to be described by a Simple Modified Lorentzian (SMLO)~\cite{PhysRevC.99.014303}:
\begin{equation}
\begin{split}
    f_\text{SMLO}(E_\gamma) = \frac{1}{3\pi^2 \hbar^2 c^2} \frac{\sigma_\text{SMLO}}{1 - \exp\left(-E_\gamma/T\right)} \\
    \times \frac{2}{\pi} \frac{E_\gamma \Gamma(E_\gamma, T)}{\left(E_\gamma^2 - E_\text{SMLO}^2\right)^2 + E_\gamma^2 \Gamma^2 \left(E_\gamma, T \right)},
\end{split} \label{eq:SMLO_func}
\end{equation}
where $E_\text{SMLO}$ is the centroid of the GDR and $\sigma_\text{SMLO}$ is the total strength of the GDR. The temperature $T$ is the temperature at final excitation energy and uses the temperature from the NLD model. The width of the SMLO is
\begin{equation}
    \Gamma(E_\gamma, T) = \frac{\Gamma_\text{SMLO}}{E_\text{SMLO}}\left(E_\gamma + \frac{(2\pi T)^2}{E_\text{GDR}}\right),
\end{equation}
and depends on final temperature and $\Gamma_\text{SMLO}$ corresponds to the width found in photo-excitations. The Pygmy-like structure found at around $7.5$ MeV is assumed to have a Gaussian shape
\begin{equation}
    f_\text{PDR}(E_\gamma) = A_\text{PDR}\exp{\left({-\frac{\left(E_\gamma - E_\text{PDR}\right)^2}{2\sigma_\text{PDR}^2}}\right)},
\end{equation}
where $E_\text{PDR}$ and $\sigma_\text{PDR}$ is the centroid and width, respectively. $A_\text{PDR}$ is the magnitude of the resonance. The choice of a Gaussian shape is due to the better fit to the data, as has also been seen in the $\mathrm{Sn}$ isotopes \cite{markova_systematics_2025}. The total $E1$ strength is
\begin{equation}\label{eq:fE1}
    f_{E1}(E_\gamma) = f_\text{SMLO}(E_\gamma) + f_\text{PDR}(E_\gamma).
\end{equation}
The $M1$ strength is assumed to consist of a spin-flip resonance and a low energy enhancement:
\begin{equation}\label{eq:fM1}
    f_{M1}(E_\gamma) = f_\text{SF}(E_\gamma) + f_\text{LEE}(E_\gamma).
\end{equation}
The spin-flip resonance is assumed to have a Lorentzian shape \cite{PhysRevC.99.014303}
\begin{equation}
    f_\text{SF}(E_\gamma) = \frac{\sigma_\text{SF}}{3\pi^2 \hbar^2 c^2} \frac{E_\gamma \Gamma_\text{SF}^2}{\left(E_\gamma^2-E_\text{SF}^2\right)^2 + \left(E_\gamma \Gamma_\text{SLO}\right)^2}, \label{eq:SLO}
\end{equation}
where $E_\text{SF}$ is the centroid, $\Gamma_\text{SF}$ is the width, and $\sigma_\text{SF}$ is the magnitude.
The LEE has a pure exponential shape
\begin{equation}
    f_\text{LEE}(E_\gamma) = C\exp\left(-\eta E_\gamma\right),
\end{equation}
where $C$ and $\eta$ have no particular physical significance.

The parameters are inferred through Bayesian inference with the likelihood
\begin{equation} \label{eq:LikeGSF}
    \begin{split}
    \mathcal{L}\left(\left\{f_\text{Oslo}\right\},\left\{f_{\left(\gamma,\gamma\right),\text{E1}}\right\},\left\{f_{\left(\gamma,\gamma\right),\text{M1}}\right\},\left\{f_{\left(\gamma,\text{tot}\right)}\right\}|\bm{\theta}\right) = \\
    \mathcal{L}_\text{Oslo} \left(\left\{f_\text{Oslo}\right\}| \bm{\theta}\right)\mathcal{L}_{E1}\left(\left\{f_{\left(\gamma,\gamma\right),\text{E1}}\right\}|\bm{\theta}\right) \\
    \times \mathcal{L}_{M1}\left(\left\{f_{\left(\gamma,\gamma\right),\text{M1}}\right\}|\bm{\theta}\right)\mathcal{L}_\text{tot}\left(\left\{f_{\left(\gamma,\text{tot}\right)}\right\}|\bm{\theta}\right)
    \end{split}
\end{equation}
where $\left\{f_\text{Oslo}\right\}$ is the set of normalized measured $\gamma$SF points, while $\left\{f_{(\gamma,\gamma),\text{E1}}\right\}$ and $\left\{f_{(\gamma,\gamma),\text{M1}}\right\}$ is the set of $E1$
 and $M1$ strength in $^{134}\mathrm{Xe}$ measured by Ref. \cite{massarczyk_magnetic_2014}. Lastly $\left\{f_{(\gamma,\text{tot})}\right\}$ is the strength function found from the total $\gamma$-absorption measurements on $^{133}\mathrm{Cs}$ from Refs. \cite{Berman1969, Lepretre1974}. Experimental uncertainties are assumed to be Gaussian, leading to likelihood function
 \begin{equation}
    \begin{split}
        \mathcal{L}\left(\left\{f_X\right\}|\bm{\theta}\right) =& \sum_j \frac{1}{2\pi\sigma_{f_{X}}\left(E_{\gamma_j}\right)} \\
        -& \frac{1}{2}\sum_j\left(\frac{f_{X}(E_{\gamma_j}) - f_Y\left(E_{\gamma_j}\right)}{\sigma_{f_X}\left(E_{\gamma_j}\right)}\right)^2
    \end{split}
 \end{equation}
where $$X=(\text{Oslo}, E1, M1, \text{tot}),$$ and $$Y=(\text{tot}, E1, M1, \text{tot}).$$ Here, $\sigma_{f_{X}}(E_{\gamma})$ is the experimental uncertainty and $\bm{\theta} = (E_\text{SMLO}, \Gamma_\text{SMLO}, \sigma_\text{SMLO}, T, E_\text{PDR}, \Gamma_\text{PDR}, A_\text{PDR}, E_\text{SF}, \Gamma_\text{SF}, \\\sigma_\text{SF}, C, \eta)$ are the model parameters.

\begin{figure*}[htp]
    \centering
    \includegraphics[width=\linewidth]{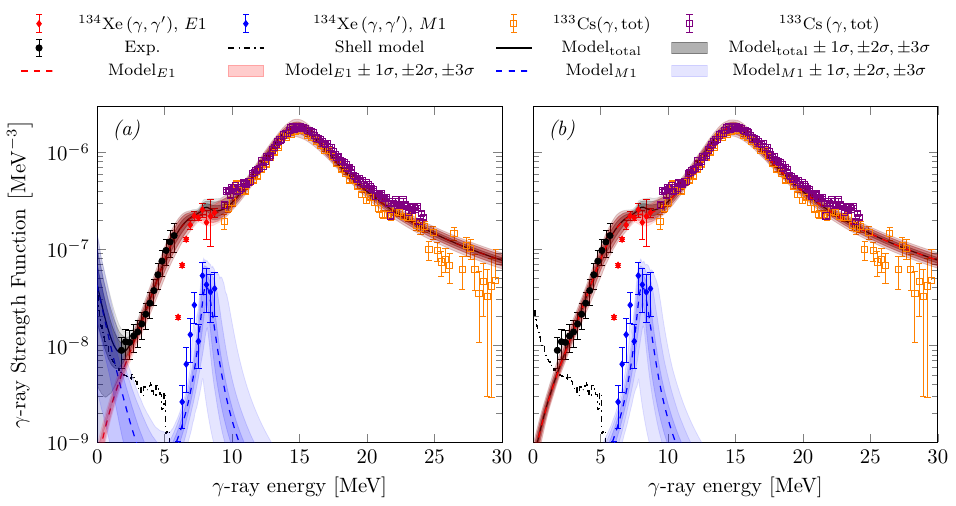}
    \caption{The \gsf of \xetres{} (filled black circles), with the credibility bands $\pm1\sigma$, $\pm2\sigma$ and $\pm3\sigma$ for the $\gamma$SF model fit. The blue and red diamonds shows the $E1$ and $M1$ $\gamma$SF of $^{134}\mathrm{Xe}$ from NRF, respectively \cite{massarczyk_magnetic_2014}. Photo nuclear data on $^{133}\mathrm{Cs}$ data \cite{Berman1969, Lepretre1974} are shown in filled purple and orange squares. The $M1$ strength function from large--scale Shell model calculation is included as a solid black line.} 
    \label{fig:gsf}
\end{figure*}

Gaussian priors are assumed of all the model parameters, with centroids and uncertainty taken from the parameterization in Ref. \cite{PhysRevC.99.014303}. In the case of $E_\text{SF}, \Gamma_\text{SF}, \sigma_\text{SF}, C, \eta$ no uncertainties are given, and are assumed to be $10\%$ for the SF and equal to the centroid for $C$ and $\eta$. The centroid for the priors of the parameters for the PDR-like structure are taken as the values that maximizes Eq. \eqref{eq:LikeGSF}, with a $50\%$ uncertainty. Lastly, the prior distribution for temperature at the final level, $T$, is the marginal posterior distribution of the CT temperature found in the normalization of the NLD and $\gamma$SF. All prior parameters are listed in Table \ref{tab:GSF_prior}. The resulting $\gamma$SF model is shown in Fig. \ref{fig:gsf}(a). As the presented results cannot determine, nor exclude the presents of a LEE, a model excluding the LEE term in Eq. \eqref{eq:fM1} is shown in Fig. \ref{fig:gsf}(b).

\begin{table}
    \centering
    \caption{\label{tab:GSF_prior}List of priors for $\gamma$SF parameterization.}
    \begin{tabular}{c|cc} \hline \hline
     Parameter & $\mu$ & $\sigma$  \\ \hline
     $E_\text{GDR}$ & $15.09$ MeV & $0.15$ MeV \\
     $\Gamma_\text{GDR}$ & $5.681$ MeV & $2.35$ MeV \\
     $\sigma_\text{GDR}$ & $2300$ mb & $230$ mb \\ \hline
     $E_\text{PDR}$ & $8$ MeV & $1$ MeV \\
     $\Gamma_\text{PDR}$ & $1$ MeV & $1$ MeV \\
     $A_\text{PDR}$ & $1.5\times10^{-7}$ MeV$^{-3}$ & $1\times10^{-8}$ MeV$^{-3}$ \\ \hline
     $E_\text{SF}$ & $8$ MeV & $0.5$ MeV \\
     $\Gamma_\text{SF}$ & $1$ MeV & $0.5$ MeV \\
     $\sigma_\text{SF}$ & $5$ mb & $1$ mb \\ \hline
     $C$ & $3.5 \times 10^{-8}$ MeV$^{-3}$ & $3.5 \times 10^{-8}$ MeV$^{-3}$ \\
     $\eta$ & $0.8$ MeV$^{-1}$ & $0.8$ MeV$^{-1}$ \\ \hline
    \end{tabular}
\end{table}

The mean of the marginal posterior distributions of the model parameters are listed in Table \ref{tab:OM_param_post}, showing the $\gamma$SF model parameters with or without a LEE term included.

\begin{table}[htbp]
	\centering
	\caption{Posterior mean of the $\gamma$SF model parameters.}\label{tab:OM_param_post}
	\begin{tabular}{c|c|c} \hline \hline
		Parameter & Value & Value (w/o LEE) \\ \hline
        $E_\text{GDR}$ (MeV) & $15.154(92)$ & $15.158(94)$  \\
        $\Gamma_\text{GDR}$ (MeV) & $4.85(34)$  & $4.86(34)$  \\
        $\sigma_\text{GDR}$ (MeV) & $2498(86)$   & $2499(88)$ \\
        $T$ (MeV) & $0.640(31)$ & $0.643(32)$ \\ \hline
        $E_\text{PDR}$ (MeV) & $7.27(53)$  & $7.33(58)$ \\
        $\Gamma_\text{PDR}$ (MeV) & $1.48(29)$ & $1.54(33)$  \\
        $A_\text{PDR}$ $\left(\text{MeV}^{-3}\right)$ & $1.421(90)\times10^{-7}$  & $1.413(89)\times10^{-7}$  \\ \hline
        $E_\text{SF}$ (MeV) & $8.15(22)$ & $8.16(22)$ \\
        $\Gamma_\text{SF}$ (MeV) & $0.64(13)$ & $0.66(13)$  \\
        $\sigma_\text{SF}$ (mb)& $4.90(89)$ & $4.90(87)$  \\ \hline
        $C$ $\left(\text{MeV}^{-3}\right)$ & $4.2(25)\times10^{-8}$  & -  \\
        $\eta$ $\left(\text{MeV}^{-1}\right)$ & $1.43(43)$  & - \\ \hline \hline
        Evidence (logZ) & $-27.51(27)$ & $-28.11(21)$ \\ \hline
	\end{tabular}
\end{table}

\section{Shell Model Calculations And Discussion}
\label{sec:shellmodel}  

\begin{figure}[htbp]
    \centering
    \includegraphics[width=\linewidth]{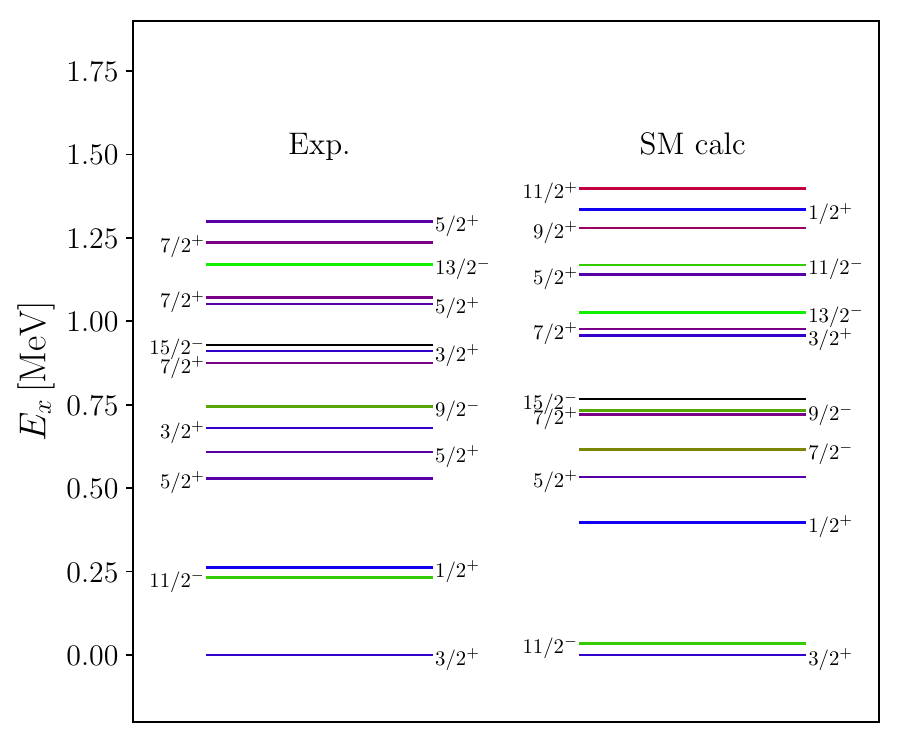}
    \caption{Comparison of the low--lying level structure from large--scale shell model calculations and measurements~\cite{CAPOTE20093107} of \xetres .} 
    \label{fig:low_lying_levels}
\end{figure}

The $M1$ $\gamma$SF and the NLD of \xetre was investigated through large--scale shell model calculation with the M--scheme shell--model code \verb+KSHELL+ \cite{KSHELL}. 
The \verb+SN100+ interaction \cite{Brown2005} with the $^{100}\mathrm{Sn}$ core model space of $(0g_{7/2}, 1d_{5/2},1d_{3/2},2s_{1/2},1h_{11/2})$ for the protons and neutrons was used. 
Four valence protons and $29$ neutrons with no truncations or limitations on occupancy within the valence space were considered. A total of $200$ eigenstates were calculated for each spin and parity of $J^\pi$ between $J=1/2$ and $J=21/2$. This results in $\approx1.2\times10^6$ $M1$-transitions with non--zero transition strength.
The low-energy level scheme of \xetre \cite{CAPOTE20093107}  is reproduced by the calculations up to $E_x \sim 1.2$ MeV as shown in Fig.~\ref{fig:low_lying_levels}.

Similarly, when using shell model calculations to model the $^{132,133}$Xe isotopes in Ref. \cite{Vogt2017}, they used the \verb+SN100PN+ \cite{Brown2005} and the \verb+PQM130+ \cite{Teruya2015} effective interactions  when comparing to their $\gamma$-spectroscopy data. From their results, both interactions reproduced the level scheme well, which is also the case with the \verb+SN100+ interaction. To investigate the effect of interaction, a similar calculation with the \verb+SN100PN+ interaction was also performed, however it overestimated the NLD, as seen by the dashed line in Fig. \ref{fig:nld}.

\subsection{Nuclear Level Density}

\begin{figure}[htbp]
    \centering
    \includegraphics[width=\linewidth]{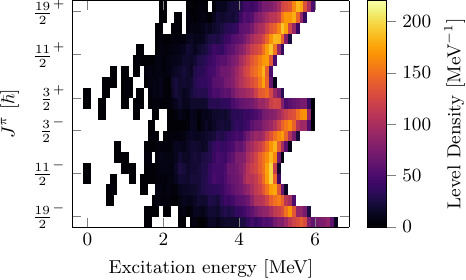}
    \caption{Decomposition of shell model NLD as $\rho(E_x,J,\pi)$, where  we see the number of levels for individual $J^\pi$ from large-scale shell model calculations up to a maximum of 200 levels for each $J^\pi$.}
    \label{fig:smnld}
\end{figure}

\begin{figure}
    \centering
    \includegraphics[width=\linewidth]{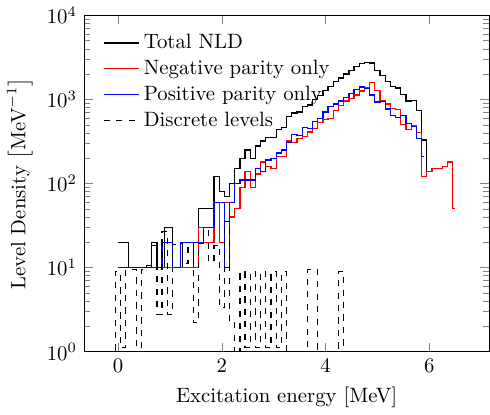}
    \caption{Calculated total nuclear level densities  (black line), negative-parity level density (red line), and positive-parity level density (blue line) compared to the experimentally available level density (dashed black line) \cite{CAPOTE20093107}.}
    \label{fig:smnld_tot}
\end{figure}

 The calculated NLD was obtained with the same binning ($300$ keV) as that used for the experimental data. The calculated levels are complete up to $E_x \sim 4.8$ MeV, where  $200$ levels with $J^\pi = 7/2^+$ have been reached. The $J^\pi$ dependent NLD is shown in Fig. \ref{fig:smnld}, while the total level density is shown in Fig. \ref{fig:smnld_tot}. 
 The $1h_{11/2}$ orbital is the only orbital with negative parity that is included in the model space, and only excitations to this orbital can generate the negative-parity states. This may be the reason of the spin distribution of negative-parity levels skewing towards higher spins in Fig.~\ref{fig:smnld}. This discrepancy may disappear with the inclusion of orbitals from the next major shell such as the $2f_{7/2}$, $3p_{3/2}$ orbits and proton orbitals within the underlying $^{100}\mathrm{Sn}$ core.
 
 The results of Fig.~\ref{fig:smnld_tot} indicate a slight asymmetry in the distribution between negative- and positive-parity states with the level density of latter $\approx 10\%$ higher. Comparison to the known level density, provides good agreement up to $\sim 1.4$ MeV and indicates the maximum $E_x$ where the experimentally-measured level scheme can be considered complete \cite{CAPOTE20093107}.

\subsection{\texorpdfstring{$\gamma$-ray strength function}{gamma-ray strength function}}

The $M1$ strength function of \xetre was calculated from the shell model for each initial excitation energy $E_i$, spin $J_i$, and parity $\pi_i$ with a binning of $100$ keV as for the experimental data. The $M1$ strength function, $f_{M1}\left(E_\gamma, E_i, J_i, \pi_i\right)$, is defined as \cite{Bart1973}:
\begin{equation} \label{eq:DefM1Strength}
    \begin{split}
        &f_{M1}\left(E_\gamma, E_i, J_i, \pi_i\right) = \\
        &\quad\frac{16\pi}{9(\hbar c)^3} \langle B(M1) \rangle(E_\gamma, E_i, J_i, \pi_i)\rho(E_i, J_i, \pi_i).
    \end{split}
\end{equation}

 According to the Brink--Axel hypothesis \cite{Brink1955,Axel1962} the strength function is assumed to be independent of initial energy, spin and parity. The resulting $f_{M1}\left(E_\gamma, E_i, J_i, \pi_i\right)$ is shown in Fig.~\ref{fig:gsf} and  low-energy enhancement at low $\gamma$-ray energies consistent with theoretical results for other nuclei \cite{Schwengner2013, Brown2014, Schwengner2017, Sieja2017, Midtbo2018, Schwengner2022, Chen2025}.

The \gsf varies slowly with nucleon number; as such, these data provide a continuous representation of the \gsf in the range $2\,\mathrm{MeV} < f\left(E_\gamma\right) < 30\,\mathrm{MeV}$. The intersections of data from the different experimental techniques are in reasonable agreement. At the lowest energies, the \gsf does not appear to continue decreasing; rather, the change in slope may indicate the onset of a low-‑energy enhancement, the manifestation of the scissors rotational mode, which is expected to be a universal feature of weakly deformed nuclei \cite{Chen2025}. The shell‑-model $M1$ strength‑-function calculations are consistent with the observed change in slope. Experimentally, it was indeed verified that this low--energy feature is of dipole \cite{Larsen2013, Larsen2017, Jones2018} and most likely of magnetic nature \cite{Jones2018}. These shell model calculations are consistent with a $M1$ nature of the low--energy enhancement.

 While $E1$ transitions were included in the calculation of Ref. \cite{Sieja2017} they are not included in the present calculations as it involves particle-–hole excitations across major oscillator shells of different principal quantum numbers. This would require a prohibitively large configuration space for the evaluation of these transition matrix elements, which would be significantly more computationally intensive. At higher energies in Fig. \ref{fig:gsf}, above our extraction region, we can compare to the $f_{\mathrm{E1+M1}}$ data from \cite{massarczyk_magnetic_2014} for $^{134}$Xe which encompasses the GDR. For the fit, we included data above 7 MeV from Ref. \cite{massarczyk_magnetic_2014}.

As can be seen for the low-energy region, the 1$\sigma$ uncertainty band of the fit includes an upbend, but for  $3\sigma$, it then encompasses the possibility of no upbend present in the $\gamma$SF. When looking at the fit decomposed to its respective parts, you can see that the LEE prediction from the shell model calculations are close to what we find within $1\sigma$. For NEEC calculations, it is this low energy region that is paramount, as the $\gamma$ energy is 5 keV in the NPI. With this fit, it will be possible to further constrain the NEEC rate based on the uncertainty bands of the \gsf, either including the predicted upbend or with no upbend.

\section{Hauser-Feshbach calculations}\label{sec:talys}
For cross sections and reaction rates, three different parts go into the Hauser-Feshbach calculations, the optical model potential (OMP), the NLD, and the \gsf. With the TALYS code \cite{koning_talys_2023}, we have calculated the $^{132}\mathrm{Xe}(n,\gamma)$-cross section with our extracted NLD and \gsfs. For each sampled set of parameters $\bm{\theta}_i$ a tabulated NLD and $\gamma$SF were produced. The tabulated NLD used the experimental NLD up to $4$ MeV, beyond the CT model was used. In the case of the $\gamma$SF separate tables for $E1$ and $M1$ were produced based on the model parameters for the $\gamma$SF model, see Eqs. \eqref{eq:fE1} and \eqref{eq:fM1}.  

\begin{figure}
    \centering
    \includegraphics[width=\linewidth]{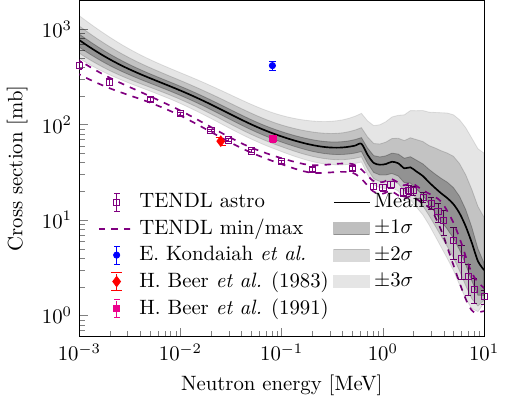}
    \caption{Constrained neutron-capture cross section on $^{132}$Xe is shown as solid black line, with gray credibility bands for $\pm1\sigma$, $\pm2\sigma$, and $\pm3\sigma$ included for this work. Recommended TENDL values \cite{ROCHMAN2025_TENDL} as purple squares. Experimental data from H. Beer \textit{et al.} (1983) \cite{Beer1983} is the filled red diamond, H. Beer \textit{et al.} (1991) \cite{Beer1991} is the filled magenta square, and E. Kondaiah \textit{et al.} is the filled blue circle.}
    \label{fig:crossSection}
\end{figure}

\begin{figure}
    \centering
    \includegraphics[width=\linewidth]{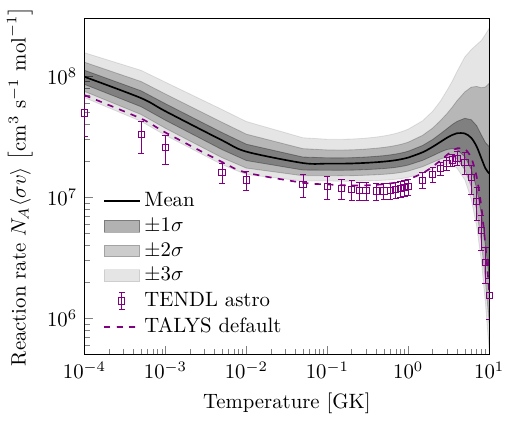}
    \caption{Constrained reaction rate for the \nreaction{} is shown as solid black line, with gray bands for $\pm1,2,3\sigma$ credibility intervals, for this work. The TALYS default is included as a dashed black line, with the minima and maxima plotted as a pink band.}
    \label{fig:talysRate}
\end{figure}

For each set of NLD and $\gamma$SF tables TALYS was run and the posterior probability distribution for the cross section and the  reaction rate is shown in Figs. \ref{fig:crossSection} and  \ref{fig:talysRate}. For the constrained neutron-capture cross section on \xeto, we compare to the recommended values from the TENDL nuclear data set \cite{ROCHMAN2025_TENDL}, where TENDL shows the predicted cross section is below our data, and below the measured ($n,\gamma$)-cross sections in literature \cite{Kondaiah1968,Beer1983,Beer1991}. Below 1 MeV, the discrepancy is much larger, wile the predicted cross section from TENDL is closer to this work above $\sim 2$ MeV.

The stellar reaction rate as a function of temperature is shown in Fig. \ref{fig:talysRate}. As seen in the figure, the reaction rate is significantly constrained by our measured NLD and $\gamma$SF. In comparison, the TENDL rate under predicts the rate below 1 GK with the recommended models in TALYS, while above 1 GK it closer to this work, and lower than the TALYS default rate.

\section{Conclusion}\label{sec:conc} 

In this paper we have presented the NLD and \gsf for \xetre which have never before been measured. This research is informative for understanding statistical properties of xenon isotopes below the neutron separation energy and can also be relevant to the NEEC process, as the \gsf is used to calculate the rate at which it is observed at low energies. These results were presented in Figures~\ref{fig:nld} and \ref{fig:gsf}.

Having both the experimental data to find NLD and \gsf made it possible to calculate the (n,$\gamma$)-cross section, Fig. \ref{fig:crossSection}, and reaction rate in Fig. \ref{fig:talysRate}. This provides a better estimate for the cross section than the default model input provided in Hauser-Feshbach codes like TALYS, or the recommended TENDL values.

\section*{Acknowledgments}
The authors would like to thank iThemba LABS operations for stable running conditions, and for the targets prepared. In addition, the authors would like to thank Ann-Cecilie Larsen for useful advice and discussion and Jon Dahl for technical help for setup of KShell. This study has been funded by the Research Council of Norway through its grants to the Norwegian Nuclear Research Centre (Project No. 341985). Additional research project grants were provided by the Research Council of Norway (Grants No. 263030 (V.W.I., S.S., A.G., F.Z., L.C.C.), 325714 (V.W.I, S.S.), and 245882 (P.J., T.L.C., F.B.G.)). H. C. Berg was provided support from the National Nuclear Security Administration under Award No. DE-NA0003180 and the Stewardship Science Academic Alliances program through DOE Awards No DOE-DE-NA0003906. This material is based upon work supported by the U.S. Department of Energy, Office of Science, Office of Nuclear Physics under Contract Nos. DE-AC02-05CH11231, DE-SC0023633, and the U.S. Nuclear Data Program (LBNL), and DE-AC52-07NA27344 (LLNL). This work is based on the research supported by National Research Foundation of South Africa (127116,118846) and iThemba LABS.

\bibliography{art.bib}

\end{document}